\documentclass[preprint,authoryear,12pt]{elsarticle}
\usepackage{epsfig}
\usepackage{amssymb}
\usepackage[
a4paper=true,%
breaklinks=true,%
colorlinks=true,%
pdfauthor={First Author et al.},%
pdftitle={Template for manuscripts in Advances in Space Research}%
]{hyperref}
\def\mnras{Mon. Not. Royal Astr. Soc.\/}
\def\apj{Astroph. J.\/}
\def\apjl{Astroph. J. Lett.\/}
\def\aj{Astron. J.\/}
\def\apjs{Astroph. J. Suppl.\/}
\def\aap{Astron. Astroph. \/}
\def\pasj{Publ. Astron. Soc. Japan}
\journal{Advances in Space Research}

\begin{document}

\begin{frontmatter}



\title{Quasars and their emission lines \\ as cosmological probes\tnoteref{footnote1}}
\tnotetext[footnote1]{Invited talk presented on May 14, 2013 at the 9th SCSLSA held in Banja Kovilia\v{c}a, Serbia.}


\author{Paola Marziani\corref{cor}\fnref{footnote2}}
\address{INAF, Osservatorio Astronomico di Padova, Vicolo dell' Osservatorio 5, 35122 Padova, Italia}
\cortext[cor]{Corresponding author}
\ead{paola.marziani@oapd.inaf.it}


\author{Jack  W. Sulentic}
\address{Instituto de Astrof{\'\i}sica de Andaluc{\'\i}a (CSIC), Glorieta de Astronom\'{\i}a, 18008 Granada, Espa\~na}
\ead{sulentic@iaa.es}


\begin{abstract}

Quasars are the most luminous sources in the Universe. They are currently observed  out to redshift $ z \approx 7$ when the Universe was less than one tenth of its present age. Since their discovery 50 years ago astronomers have dreamed of using them as standard candles.  Unfortunately quasars cover a very large  range (8 dex) of luminosity making them far from standard.  We briefly review several methods  that can potentially exploit quasars properties and allow us to obtain useful constraints on principal cosmological parameters. Using our 4D Eigenvector 1 formalism we have found a way to effectively isolate quasars radiating near the Eddington limit. If the Eddington ratio is known, under several assumptions it is  possible to  derive distance independent luminosities. We discuss and the main statistical and systematic errors involved, and whether    these ``standard Eddington candles'' can be actually used to  constrain cosmological models. 
\end{abstract}

\end{frontmatter}

\parindent=0.5 cm

\section{Introduction}

A majority of quasars can be identified from their distinctive emission line spectra and continua rising toward the UV.  The emission lines are broad (FWHM $\gtrsim$ 1000 km ~s$^{-1}$\ ) and strong (equivalent width $\sim 10^2$ \AA\ in the rest frame). Quasars are numerous, very luminous (even $L \sim 10^{48}$ erg s$^{-1}$) and observed over a wide redshift range ($0 \lesssim z \lesssim  7$). We can only repeat a question that we asked  years ago \citep{marzianietal03d}: why have quasars never been successfully used as cosmological probes? 

The reasons that quasars have not been used for cosmology (i.e. like supernov\ae) remains the same as ten years ago: they are sources with an  evolving  luminosity function that is open-ended at low $L$\ \citep[e.g.][]{grazianetal00,boyleetal00,palanqueetal13}. 
Spectral properties  do not show strong signs of a luminosity dependence or, better said, quasars do  show  
self-similar broad line spectra with varying luminosity, black hole mass, and redshift. This is not to say that all quasar spectra are identical but  that neither luminosity nor  mass are the main drivers of  quasar spectral diversity.  Quasars are also  anisotropic sources most easily seen at radio frequencies where relativistic beaming is observed in some  radio-loud sources  \citep[e.g.,][]{urrypadovani95}.

Given these properties it is reasonable to ask if quasars can tell us anything about the geometry of the Universe. 
After fifty years of study the answer to this question appears to be a cautious ``yes.''  A first hint comes from consideration of the Hubble diagram for the brightest quasars. One can predict the apparent magnitude of a quasar with
``maximum'' mass and radiating at Eddington limit and derive some constraint on the distance scale. 
If this computation is carried out \citep{bartelmannetal09}, one  derives $H_{0}$\ values that  are in agreement with present-day estimates ($\approx$ 60-70 km ~s$^{-1}$\  Mpc$^{-1}$). Clearly we are a far cry from the original Hubble estimate of  500 km ~s$^{-1}$\  Mpc$^{-1}$\  \citep{hubble29}. 

\section{Major approaches}

We can go much beyond the elementary consideration outlined above. In the last decades several techniques were devised  to exploit quasars for cosmology. They can be grouped into three major approaches:

\begin{itemize}
\item correlations with luminosity;
\item identification and measurement  of standard rulers (often from time delays);
\item identification of ``Eddington standard candles'' in the general quasar population. 
\end{itemize}

Other methods are not based on the intrinsic properties of quasars but rather on the matter 
distribution mapped through the light emitted by quasars (for example baryon acoustic oscillations 
in the quasar Ly$\alpha$ forest; \citealt{buscaetal13}) or the spatial distribution of quasars. 
They will not be further considered here.  In the following we will briefly review the merits of the 
first two approaches and focus on the ``Eddington standard candle'' approach. 

\section{Luminosity correlations}
\label{baldwin}

An anti-correlation between the equivalent width (EW) of high ionisation C{\sc iv}1549 and continuum specific 
luminosity  was found in a small sample of core-dominated radio loud sources \citep{baldwin77}.  
This ``Baldwin effect'' -- as the anti-correlation came to be known -- took on a life of its own and was 
confirmed in a large number of subsequent studies (see \citealt{sulenticetal00a}  for a synopsis up 
to mid-1999) but always with increasing looseness. Modern formulations of the Baldwin effect 
involve a weak anti-correlation (with slope $\approx$ -0.1) that becomes significant only if large 
samples are considered. Most recent  detections of the Baldwin effect consider samples in excess of 
20000 sources \citep{bianetal12}.   The cosmological expectations raised by the original Baldwin 
Effect foundered on the large dispersion observed in subsequent studies. The discovery of a sizable 
population of low-$z$, low luminosity  quasars with EW C{\sc iv} comparable to values observed in high-luminosity quasars also  made the effect sample-dependent \citep{kinneyetal90}:  low C{\sc iv} EW sources at low luminosity are mainly  Narrow Line Seyfert 1s (NLSy1s) that are more frequent in soft X-ray selected samples.  The major factor governing the Baldwin  effect was unveiled by the discovery that  EW C{\sc iv} is much more strongly dependent on Eddington ratio than on luminosity \citep{bachevetal04,baskinlaor04}:  \citet{baskinlaor04} found 
correlation coefficients $r_S \approx -0.6$ and $r_S \approx -0.15$\ for  Eddington  ratio and 
luminosity, respectively.  This result implies a strong role for selection effects: in a flux limited sample the
higher Eddington radiators are preferentially selected at larger distances. It is possible to show that 
a weak correlation between equivalent width and luminosity will also arise in an ideal complete sample 
because of the strong correlation between EW and Eddington ratio and that a flux limit makes it 
possible to account for the observed correlation slope \citep{marzianietal08}.

\section{Measuring   standard rulers in quasars}

Generally speaking the standard rulers sought for cosmological purposes are either uncomfortably extra large (weak lensing), or exceedingly small (quasars). Restricting the attention to quasars, a linear size that can be used as a standard ruler is the distance between the broad line emitting gas and the central continuum source (hereafter the broad line region radius, $r_\mathrm{BLR}$). This distance has been measured via reverberation mapping for $\approx$ 60 active nuclei and quasars at $z \lesssim 1$\ \citep{bentzetal13} with programmes underway to measure $r_\mathrm{BLR}$ in more distant quasars \citep[e.g., ][]{kaspietal07,treveseetal07,cheloucheetal12,wooetal13}. Line luminosity arises from photoionization by  an FUV continuum and
lines respond to continuum luminosity changes with a time delay, $r_\mathrm{BLR} \approx c \tau$.  The radius $r_\mathrm{BLR}$ is obtained by measuring  the  peak or centroid   displacement  of the cross-correlation function between the light curves of the continuum and a strong line. While the meaning of this measure is not fully clear \citep{devereux13}, $r_\mathrm{BLR}$ is measured in a way that is redshift independent. If it were possible to measure the angular size of the BLR then a redshift independent value of the angular distance $d_\mathrm A$\ would follow: $d_\mathrm A(H_0, \Omega_\mathrm{M}, \Omega_\Lambda) = c\tau / \theta''$\ \citep{elviskarovska02}. However, our  ability to resolve the broad line region is still beyond reach of the most advanced optical interferometers (ESO VLTI, etc.):  the angular size subtended by the broad line region in some of the nearest sources is no more than a few tenths of milliarcsecond. 

A second method employing the BLR size as a standard ruler  is based on the expected dependence of $r_\mathrm{BLR} $\	 with luminosity, originally predicted on the basis of photoionization physics \citep{davidson72,krolikmckee78} and observationally confirmed from from reverberation data: $r_\mathrm{BLR}  = c \tau \propto \sqrt{L}$\ \citep[][and references thererin; see also \citealt{kaspietal05}]{bentzetal13}.  The ratio $\tau/\sqrt{\lambda f_\lambda}$ \ is proportional to $ d_{\mathrm L} (H_0, \Omega_\mathrm{M}, \Omega_\Lambda)$\ \citep{watsonetal11,czernyetal12}. A challenge is to measure $\tau$ for a large number of objects. Chances of success may have increased using the newly devised  technique of photometric reverberation mapping \citep{haasetal11} that is less time-demanding than spectroscopic monitoring. 


\begin{figure}
\begin{center}
\includegraphics*[width=6.5cm,angle=0]{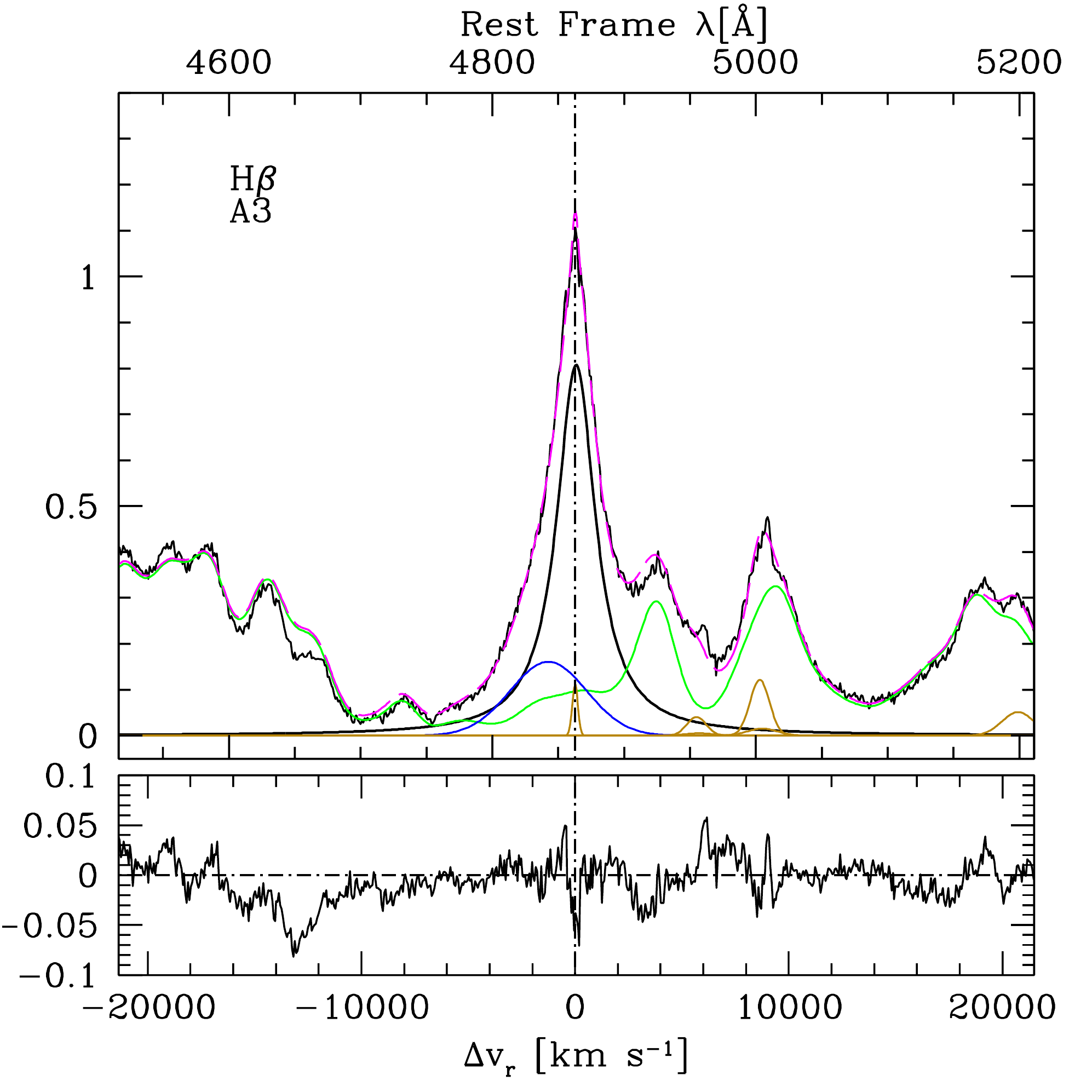}
\includegraphics*[width=6.5cm,angle=0]{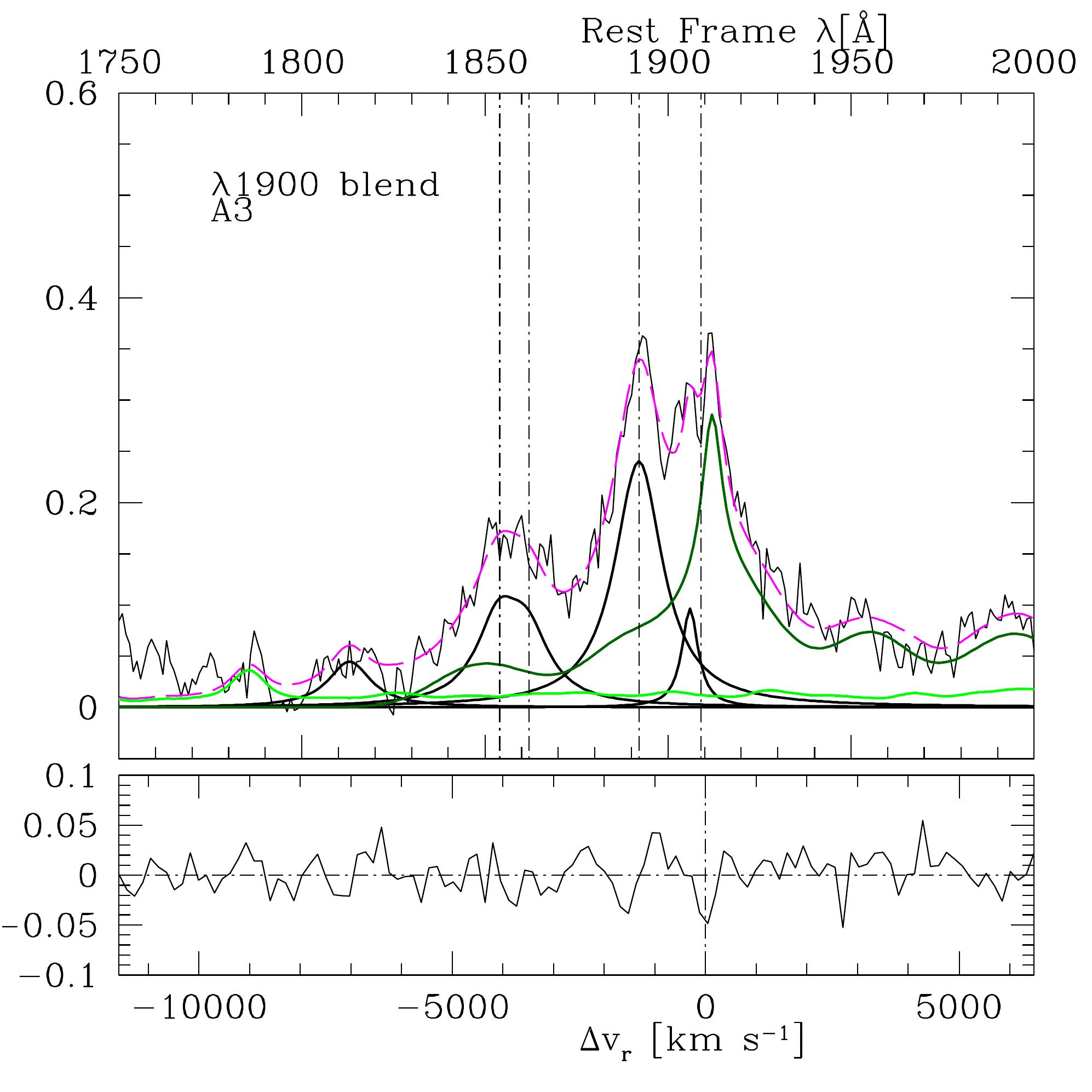}
\end{center}
\caption{Selection of high Eddington ratio candidates from emission line properties. The left panel shows the continuum subtracted spectral region around H$\beta$\ of the spectral type A3 median composite computed by \citet{marzianietal13}; the right panel  the 1900 \AA\ range for the same spectral type from  \citet{bachevetal04}. The magenta dashed lines show  models of the total line emission. Other lines trace individual line components or emission templates: broad components (thick   filled), narrow line components  (orange), Fe{\sc ii} templates  (pale green), Fe{\sc iii} template in the UV  (dark green), blue shifted component of H$\beta$\ (blue). The dot-dashed vertical lines identity the rest frame wavelength of H$\beta$\ (left panel) and the doublet Al{\sc iii}$\lambda$1860, Si{\sc iii}]$\lambda$1892, and C{\sc iii}]$\lambda$1909 (right). The intensity C{\sc iii}]$\lambda$1909 is poorly determined because of the prominent Fe{\sc iii} emission and possible Ly$\alpha$\ pumping of Fe{\sc iii}$\lambda$1914. Even if no  Ly$\alpha$\ pumping is assumed the intensity of  C{\sc iii}]$\lambda$1909 remains lower or similar to  the one of Si{\sc iii}]$\lambda$1892.   \label{fig:a3}
}
\end{figure}

\section{Eddington standard candles}

\subsection{The 4D Eigenvector 1 space}

It is now widely accepted that quasars are powered by accretion onto a massive compact object -- 
most likely a supermassive black hole \citep[see e.g.,][and references therein]{donofrioetal12}. Mass accretion rate governs the bolometric  luminosity $L$; however, observed spectroscopic parameters are too weakly dependent on $L$\ to 
provide a redshift independent estimate of bolometric luminosity.  Quasars are an almost self similar phenomenon 
over an extremely wide range of black hole mass \citep{zamanovmarziani02}. Nonetheless, in the course of the last two decades it has been possible to organize quasar diversity along systematic trends. The first successful attempt involved a principal component analysis of measures for bright PG quasars \citep{borosongreen92}. Subsequent studies revealed correlations between the first eigenvector (E1) of quasar samples and other observational parameters  \cite[e.g.,][]{wangetal96,sulenticetal00a,sulenticetal07,kruzceketal11,tangetal12}. Source luminosity (associated with the second eigenvector) and high-ionization emission line properties are more weakly correlated \citep{dietrichetal02}, as   summarized in \S\ \ref{baldwin}.  

A 4D Eigenvector 1 parameter space using four measures has proved especially useful for characterising E1 trends
\citep{sulenticetal00a,sulenticetal00b,sulenticetal07}. The measures are associated with: 1)
velocity dispersion in the low-ionization line emitting region i.e., the FWHM of H$\beta$, 2) physical conditions in the broad line region, described by the intensity ratio between the {Fe}{\sc ii}$_\mathrm{opt}$\  blend centered at 4570 \AA\ and H$\beta$, $R_\mathrm{Fe} = I($Fe{\sc ii}$\lambda 4570)/I(H\beta$), 3) strength of a soft X-ray thermal continuum component represented by the soft-X photon index, and 4) amplitude of systematic nonvirial motions in high ionization gas i.e., the amplitude of the C{\sc iv}$\lambda$ 1549 line blueshift.  Eddington ratio is likely to be the principal driver of 4DE1   \citep[e.g.,][]{borosongreen92,marzianietal01,marzianietal03b,baskinlaor05b}.
The elongated source distribution  in the optical plane of 4DE1 defined by
FWHM H$\beta$\ vs. $R_\mathrm{Fe}$, appears to be an Eddington ratio sequence \citep[see e.g.,][for diagrams showing the source distribution for large low-$z$\
samples]{marzianisulentic12a,zamfiretal10}. As the word say, E1 is intrinsically a 1D vector in an $n$-dimensional space of parameters. There are several   parameters that correlate with the original E1 in addition to the four ones included in the 4DE1 formulation. Among them, the hard X--ray continuum shape \citep[e.g., ][]{fanalietal13}, and the prominence of Al{\sc iii}$\lambda$1860 emission \citep{bachevetal04,negreteetal12}. Could one of more E1 correlates be the path to selecting standard Eddington candles?

\subsection{Extreme Eddington radiators}

If the Eddington ratio is known ($L/L_\mathrm{Edd} \propto L/M_\mathrm{BH}$), then the bolometric luminosity can be 
derived if the mass   $M_\mathrm{BH}$\ is also known. Estimates of black hole mass have now been made 
for tens of thousands of quasars \citep{marzianisulentic12,shen13} following the assumption that some emission 
lines are broadened by virial motion. Under the virial assumption it is possible to compute   $M_\mathrm{BH} = f_\mathrm{S} r_\mathrm{BLR} \mathrm{FWHM}^2 / G$\ (where  $f_\mathrm{S}$ is a structure factor $\approx$ 1) and hence derive the bolometric luminosity $L$. In principle this approach  can be applied to  quasars of any Eddington ratio \citep{davislaor11} but, in practice, attempts have been focused on a minority of quasars that are believed to radiate close to  an extreme luminosity associated with the Eddington limit.  

The condition $L/L_\mathrm{Edd} \rightarrow$ 1 (up to a few times the Eddington luminosity)  is   physically motivated.  When the mass accretion rate becomes super-Eddington, emitted radiation is advected toward the black hole, so that the source luminosity   increases only with the logarithm of accretion rate  \citep{abramowiczetal88,mineshigeetal00}. The accretion flow remains optically thick 
 so that radiation pressure  ``fattens'' it. 
 
\subsubsection{The steepest X-ray sources} 
 
 The resulting ``slim'' accretion disk is expected to emit a steep soft and hard X-ray spectrum,  with hard X-ray photon index (computed between 2 and 20 KeV) converging toward $\Gamma_\mathrm{hard} \approx 2.5$ and bolometric  luminosity saturating to 

\begin{equation}
L \approx \lambda_{\rm L} \left[ 1 + const.  \ln \left(\frac{\dot{m}}{50}\right)\right] M_\mathrm{BH},
\end{equation}
where $\dot{m}$\ is the dimensionless accretion rate \citep{mineshigeetal00}, and $\lambda_{\rm L}$\ is a constant related to the asymptotic $L/M_\mathrm{BH}$\ ratio for $\dot{m} \rightarrow \infty$.  This result, along with the expression for virial black hole mass,  allowed  \citet{wangetal13} to write a redshift independent formula for the quasar bolometric luminosity.  Magnitude differences between the $z$ independent estimates and the standard estimates based on redshift converge to 0 with a scatter that is $\approx 1.$ magnitude at 1 $\sigma$\ confidence level, if $2.3 \lesssim \Gamma_\mathrm{hard} \lesssim 2.5$.  This method is based on the theoretical prediction of the existence of super-Eddington accretors whose hard X-ray spectrum shows a steep slope. 

A  challenge in this case is to find a sample that is large enough because only 12 suitable objects at $0 \lesssim z \lesssim 0.5$\ have been found by \citet{wangetal13}. In addition, we did not favor  hard X-ray measures as key 4DE1 parameters \citep{sulenticetal00a} because they showed weaker correlation with optical/UV line parameters than $\Gamma_\mathrm{soft}$.  Current X-ray databases do not allow us to exploit the soft X-ray excess as a selector of Eddington candles. Instead we look to other UV parameters that are more closely correlated with 4DE1 parameters most tightly dependent on Eddington ratio.  

\subsubsection{The strongest Fe{\sc ii} emitters}
\label{strongestfeii}

Extreme sources (hereafter extreme Pop. A sources in 4DE1: or xA) do not show only a soft X-ray excess or a steep X-ray continuum but also the largest $R_\mathrm{Fe}$ values. This measure is already available for hundreds of low $z$\ quasars. 
A potential  4DE1 correlate involves the prominence of the  resonance doublet of Al{\sc iii}$\lambda$1860 \citep{sulenticetal07,negreteetal12}. The appearance of the optical and UV spectrum of these sources is shown in Fig. \ref{fig:a3}.  Avoiding any line width definition, the following two criteria are suitable for selecting high-$L/L_\mathrm{Edd}$\ candidates over a wide $z$\ range: 1)  $R_\mathrm{Fe} \gtrsim 1.0$; 2) $I$(Al{\sc iii}$\lambda$1860) $\gtrsim$ 0.5 $I$(Si{\sc iii}]$\lambda$1892) (Marziani \& Sulentic, submitted). These conditions  are sufficient to isolate sources radiating close, or better said, {\em closest} to the Eddingon limit. They are satisfied by $\approx$ 10\%\ of a low-$z$\ sample based on the SDSS  \citep{zamfiretal10}.  The width of Al{\sc iii}$\lambda$1860, Si{\sc iii}]$\lambda$1892, and H$\beta$ are 
extremely well correlated \citep{negreteetal13}.
  
Under the virial assumption, it is possible to write the bolometric luminosity of a source radiating at a given Eddington ratio in terms of the line broadening:

\begin{equation}
L(\delta v)  \approx   \left( \frac{\zeta_\mathrm{L}^{2}}{4\pi c h G^{2}}\right) f_\mathrm{S}^{2} \left(\frac{L}{L_\mathrm{Edd}}\right)^{2} \left(\frac{  \kappa}{\bar{\nu_\mathrm{i}}}\right)  \frac{1}{(n_\mathrm{H}U)} (\delta v)^{4},   \label{eq:vir}
\end{equation}

where $\zeta_\mathrm{L} \approx 10^{4.81}$\ erg s$^{-1}$\ g$^{-1}$, the ionizing luminosity is assumed to be $L_\mathrm{ion}=\kappa L$, with  $\kappa \approx 0.5$. The variable $\bar{\nu_\mathrm{i}}$\ is the average frequency of the ionizing photons, and the product $n_\mathrm{H}U$, density times ionisation parameter, is the ionizing photon flux. Eq. \ref{eq:vir} assumes a strict validity of the relation $r_\mathrm{BLR} \propto \sqrt{L}$. This assumption appears to be verified for the general population of quasars, i.e., for quasars of any Eddington ratio, and should be even  more true for a sample of quasars whose   Eddington ratio values cluster around a fixed  value with a small dispersion. A violation   would imply luminosity-dependent effects on the diagnostic line ratios which we do not see. 

\begin{figure}
\label{figure2}
\begin{center}
\includegraphics*[width=9.cm,angle=0]{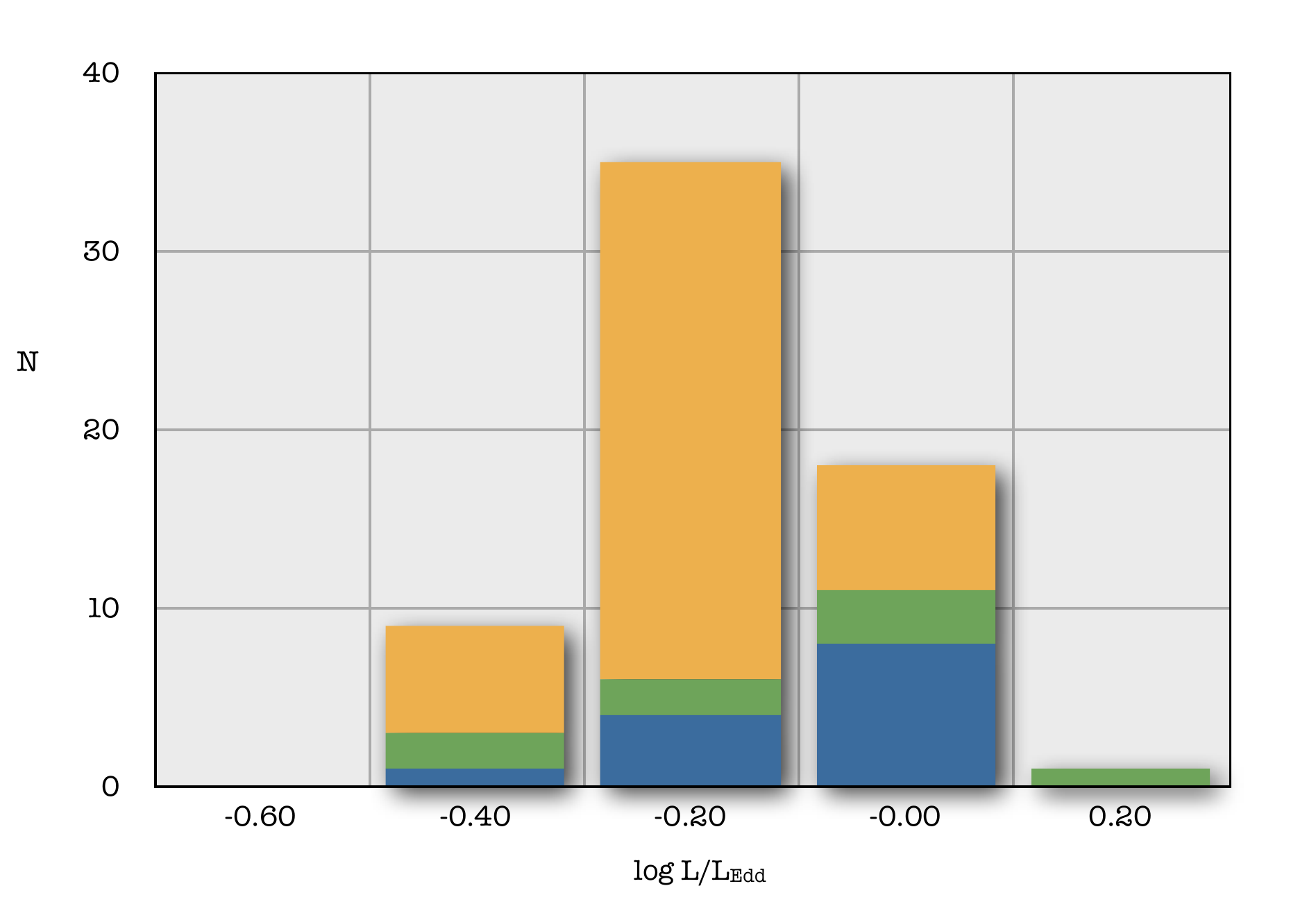}
\end{center}
\caption{A posteriori distribution of Eddington ratio for the preliminary sample of 63 high Eddington ratio sources analyzed by Marziani \& Sulentic (2013, in preparation). The different colours identify three subsamples:  sources in the redshift range $0.4 \le z \le 0.75$ \citep[green][]{marzianietal13};   sources from an Hamburg-ESO sample based on H$\beta$ IR observations \citep[blue][]{marzianietal09};    sources from an SDSS selected sample based on the appearance of the $\lambda$1900 blend (yellow).  \label{fig:edd}}
\end{figure}

Empirical estimates  show that the $L/L_\mathrm{Edd}$\ distribution decreases near $L/L_\mathrm{Edd} \approx$ 1 \citep[e.g., ][]{woourry02,shenetal11}. When the  $L/L_\mathrm{Edd}$\  distribution for xA sources is computed, they are found to be the sources closest to the Eddington limit, clustering around the largest values with small dispersion. The a-posteriori distribution of  Eddington ratio values for a sample selected according to the diagnostic  criteria indeed shows  clustering $\rightarrow$ 1 with  small dispersion, $\approx 0.15$ dex (Fig. \ref{fig:edd}).  

There is no  claim that the  distribution of Fig. \ref{fig:edd} is peaked around a {\em true} (unbiased) Eddington ratio value. Several systematic effects could influence our estimate. The most relevant one is related to the orientation of the emitting region. The expected systematic effect of orientation   increases the average $L/L_\mathrm{Edd}$: if the broad line region is flattened, then our computations will systematically overestimate $L/L_\mathrm{Edd}$ since the FWHM will be reduced in all cases because the line-of-sight component of the virial velocity will appear always less than the true one. However, this is not affecting the usefulness of xA sources as Eddington standard candles, provided that they really cluster around {\em some} value of Eddington ratio with a small dispersion.  Fig. \ref{fig:edd} shows an a-posteriori confirmation that the $L/L_\mathrm{Edd}$\ distribution is tight.  This results  does not depend on cosmology:   changing values for the $\Omega$s affects little the distribution dispersion. The $L/L_\mathrm{Edd}$\ entering Eq. \ref{eq:vir} has been computed from the $L/M_\mathrm{BH}$\ ratio, and  $L$\ assuming a value of $H_0$ at very low redshift when the $\Omega$s' influence is negligible. 

Strictly speaking, even if all other parameters are expression of the quasar intrinsic properties,  $H_0$\ cannot be independently estimated because of circularity in  assuming $L/L_\mathrm{Edd} \approx 1$. Avoiding any circularity in   Eq. \ref{eq:vir}  would require $z$-independent bolometric luminosity estimates for very low-$z$\ extreme accretors, for example from   type Ia supernov\ae\ in the host galaxies or from  reverberation mapping measurements of  time delay. In practice, the values of the quasar parameters entering Eq. \ref{eq:vir} are so uncertain  that the derived $H_0$\ is   poorly determined  and has to be assumed. The key in using Eq. \ref{eq:vir} for  $\Omega$s  estimation is that the dispersion  around any parameter  value   is small, not that we know its {\em true}  value. 

Are  xA sources the super-Eddington accretors postulated by \citet{wangetal13}? xA sources are relatively frequent at low-$z$, $\approx$ 10\%\ of optically selected samples. The estimates of $L/L_\mathrm{Edd}$ are all below $L/L_\mathrm{Edd}\approx$ 2, the value predicted for super-Eddington accretors.   Both the super-Eddington accretors of \citet{wangetal13} and xA sources are sought looking at one  extreme property. The extreme properties are correlated in the 4DE1 context, so that it is likely that super-Eddington accretors are included in the xA sample, but they may be a   fraction of all xA sources. 

\subsection{A preliminary application}

Can any  method based on Eddington ratio standard candles be applied   to actual data  and give relevant results? We selected a preliminary sample comprising sources in three redshift ranges. Of two H$\beta$ samples, one was based on  SDSS spectra with  $0.4 < z < 0.75$\ and a second one  on published H$\beta$ VLT ISAAC observations, with $0.9 < z < 1.5$.   A third sample used the criterion on Al{\sc iii}$\lambda$1860 and was based on SDSS in the redshift range  $2< z < 2.6$.  The three samples yielded 63 sources in total.  We  then compared the ``virial  luminosity'' $L(\delta v)$\ and bolometric luminosity $L$ estimated for various cosmological models.

\begin{equation}
\Delta \log L   =  \log L(\delta v)   - \log L ( z,  H_0, \Omega_\mathrm{M},
\Omega_{\Lambda})
\label{eq:delta}
\end{equation}

\begin{figure}

\begin{center}
\includegraphics*[width=9cm,angle=0]{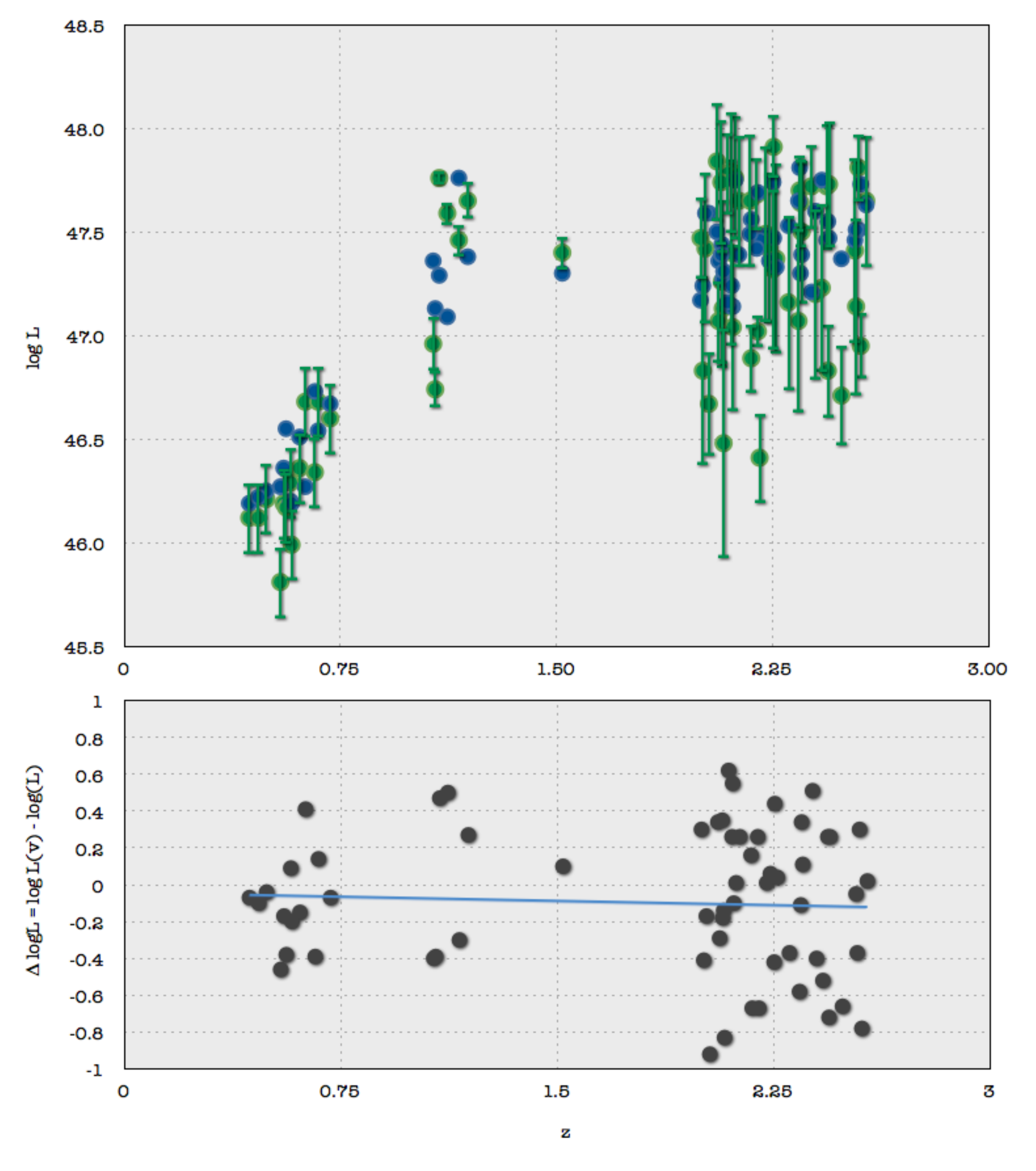}
\end{center}
\caption{Top: Decimal logarithm of bolometric luminosity $L(\delta v)$  computed from the virial equation (green), and from the customary relation involving specific flux and redshift\ (blue), assuming concordance cosmology versus redshift $L (z, H_0, \Omega_\mathrm{M}, \Omega_\Lambda)$,  for a preliminary sample of 63 quasars discussed in this paper. The bottom panel shows residuals $\Delta \log L   =  \log L(\delta v)   - \log L (z,  H_0, \Omega_\mathrm{M}, \Omega_\Lambda)$.  In principle the valid cosmological model should give 0 average and slope 0 for the residuals. 
\label{fig:cosmo}}
\end{figure}

The upper panel of Fig. \ref{fig:cosmo} shows the ``Hubble diagram'' for the preliminary sample. Virial bolometric luminosity (computed from the best guess of parameters in Eq. \ref{eq:vir}) and $z$-based bolometric  luminosity (computed for  concordance cosmology; \citealt{hinshawetal09}) are plotted as a function of redshift. The lower panel shows the residuals. Residual average is different from 0 and their slope is slightly nonzero. The differences are not significant and this preliminary application shows consistence with concordance cosmology. With the rms and the size of the preliminary sample we can statistically exclude only models like a $\Lambda$- ($\Omega_\Lambda = 1, \Omega_\mathrm{M} = 0$) or matter-dominated ($\Omega_\Lambda = 0$, $\Omega_\mathrm{M} = 1$) flat Universe. In the first case the slope of the residuals becomes strongly negative because   $L ( z,  H_0, \Omega_\mathrm{M}, \Omega_{\Lambda})$ is too large with respect to the virial luminosity; if instead $\Omega_\Lambda = 0$, $\Omega_\mathrm{M} = 1$\ is assumed,  $L ( z,  H_0, \Omega_\mathrm{M}, \Omega_{\Lambda})$ \ is too small and the slope is positive. 

\subsubsection{Uncertainties}

More than gaining conclusive results (apart from excluding extreme models already ruled out since long), experimenting with a small sample of 63 objects helped us understand the requirements for a sample of quasars able to set meaningful constraints on cosmological parameters. As mentioned, the virial bolometric luminosity   depends on several parameters whose precise value is poorly known but that should be the same for all sources radiating at limiting Eddington ratio  with a hopefully small dispersion. The H$\beta$ profile of xA sources can be almost always modelled with an unshifted (virial) Lorentzian component plus an additional component affecting the line base. The profile similarity suggests the same, reproducible structure.  The parameters related to the continuum shape should also show very small dispersion since xA sources show very similar emission line spectra \citep{nikolajuketal04}. 
 The sample standard deviation of $\Delta \log L$\ in our preliminary sample is $\approx$0.4 dex. It is a large value that reflect in parts the low S/N of most spectra. 

Looking in more detail at the parameters entering Eq. \ref{eq:vir}, we can estimate each parameter's contribution to the total error  (a more detailed discussion of the error budget is presented in  Marziani  \& Sulentic, 2013, submitted). The factor $\kappa/\bar{\nu_\mathrm{i}}$\ may appear especially uncertain since it involves the scantly observed far UV quasar spectral energy disturbution. However, we have to make a restriction to spectral energy distributions appropriate for Pop. A sources, at least, or for large $R_\mathrm{Fe}$\ NLSy1s, at best. We considered four continuum models: the ones of \citet{mathewsferland87} and \citet{koristaetal97}, a NLSy1 spectral energy distribution derived from \citet{grupeetal10} and \citet{panessaetal11}. Since $\kappa \approx .5$, and $h \bar{\nu_\mathrm{i}} \approx 40 - 60$ eV, there is a scatter of 0.033 dex in ionizing luminosity induced by the differences in the four continua. The ionizing photon flux $U n_\mathrm{H}$\  is estimated from the diagnostic ratios. An average value (10$^{9.6}$ cm$^{-3}$) has been assumed in the calculation of the virial luminosity for the sources of the preliminary sample.  Dispersion has been estimated to be $\approx$ 0.1 dex.  The structure factor $f_\mathrm{S}$\ has been derived by scaling the correlation between virial product $r_\mathrm{BLR} \delta v^2/G$\ and  host galaxy velocity dispersion to agree  with the same correlation for non-active galaxies \citep{onkenetal04}, and its  uncertainty is $\delta f_\mathrm{S}/f_\mathrm{S} \approx 0.2$.  Our sources show very similar H$\beta$\ profiles, all of them consistent with a Lorentzian component and a blue shifted additional component. Since $L/L_\mathrm{Edd}$\ is probably a major factor governing BLR structure, we may expect $\delta f_\mathrm{S}/f_\mathrm{S} \ll 0.2$. Statistical errors on the virial luminosity should be combined with errors associated to the conventional bolometric luminosity determination. Uncertainties on  $z$, spectrophotometry, and bolometric correction contribute to the error budget with  $\approx$0.1 dex. We obtain a total  rms $\approx$ 0.4, consistent with the scatter of our  preliminary data. The main hope for improving statistical error resides in more precise measures of FWHM since the line broadening enters with the fourth power in Eq. \ref{eq:vir}. Reducing FWHM measurements uncertainty from 15\% to 5\%\ would reduce the total scatter to $\approx$0.3 dex. Therefore, the rms could be lowered to  $\approx$0.3 dex by simply considering better data. 
 

The most important systematic effect that we were able to identify (Sulentic \& Marziani, submitted) is related to the bolometric correction needed to convert monochromatic luminosities at 5100 \AA\ and 1800 \AA\ into bolometric luminosity, or better, to the ratio $\lambda f_{\lambda,5100}/\lambda f_{\lambda,1800}$. Likely values of this ratio range from 0.63 (the standard \citealt{mathewsferland87} model continuum) to 0.7 (the observed I Zw 1 spectrum). A small change in the assumed $\lambda f_{\lambda,5100}/\lambda f_{\lambda,1800}$\  has a significant effect on    $\Omega_\mathrm{M}$\ because it affects the average $\Delta \log L$\ at  high redshift. Calibration observations of sources for which H$\beta$ and the 1900 blend  are simultaneously covered are needed to determine an average $\lambda f_{\lambda,5100}/\lambda f_{\lambda,1800}$\  and its dispersion.  

Systematic differences in the parameters entering Eq. \ref{eq:vir} \ as a function of the diagnostic ratio $R_\mathrm{Fe}$\ are  also expected. They will contribute  to the overall sample rms but should not introduce any systematic effects if the  distribution of $R_\mathrm{Fe}$\ values remains constant with $z$. The sample rms could be further  reduced if a trend involving $R_\mathrm{Fe}$\ and  ionizing photon flux is found. Exploring this possibility will become feasible  once a large sample of xA sources is created. %

\section{Comparison with Supernov\ae,  Prospects, and Caveats}

\subsection{Concordance cosmology from supernov\ae\ and prospects for Eddington standard candles}

Only few Supernov\ae\ at $z\gtrsim$ 1 had been discovered at the turn of the century. At that time the Hubble diagram with type Ia supernov\ae\ showed a large scatter, although the supernova data points appeared systematically fainter than expected. The observations sampled mainly the redshift interval where the effect of nonzero $\Lambda$\ yields an accelerated expansion. At the time of writing,  supernova surveys have produced and analysed data for $\approx 500$ supernov\ae\ \citep{conleyetal11}, although the wide majority are still at $z\lesssim 1$. In addition the WMAP 9 yr combined results and the Planck probe results indicate that $ \Omega_\mathrm{M}$\ and  $\Omega_{\Lambda}$\ are known with an accuracy of a few percent. Surveys are planned to improve the precision down to 1-percent level. The accelerated expansion of the Universe is perhaps not anymore an issue. 

\begin{figure}

\begin{center}
\includegraphics*[width=10cm,angle=0]{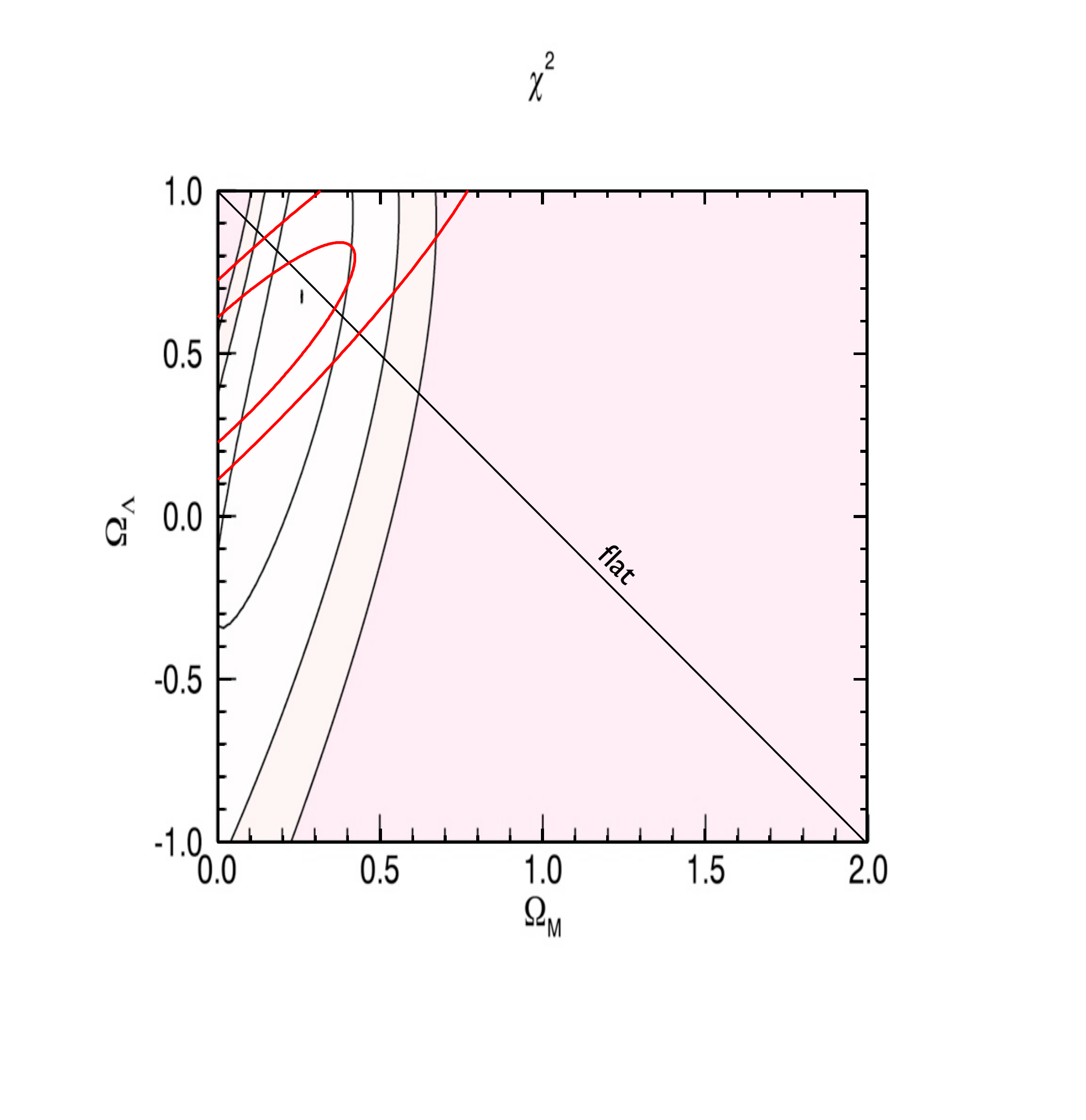}
\end{center}
\caption{Comparison between the constraints set by the supernova photometric survey described by \citet{campbelletal13} (red lines) and an hypothetical mock sample of 400  quasars with rms = 0.3 that assumes concordance cosmology  (shaded contours). Confidence intervals are at 1 and 2 $\sigma$\ for supernov\ae\ and 1, 2, 3  $\sigma$\ for the quasar mock sample. The flat geometry loci $ \Omega_\mathrm{M}+ \Omega_{\Lambda} = 1 $\  are also shown. Note the potential ability of the quasar sample to better constrain  $ \Omega_\mathrm{M}$. Only statistical errors are included in both cases. \label{fig:omegas} }
\end{figure}

There is however an advantage due to the ability of quasar data to cover almost uniformly the range between 0 and 4. Supernov\ae\ have been discovered mainly at $z \lesssim 1$, and Planck and WMAP deal after all with features detected at the surface of last scattering when the Universe became transparent to radiation, at $z \sim 1000$. Quasars can sample cosmic epochs when the negative pressure of dark energy was dominating, as well as earlier epochs when the Universe expansion  was still dominated by the effect of matter ($z \gtrsim 1$). If 400 xA sources with rms $\approx$0.3 dex can be found  then constraints on $ \Omega_\mathrm{M}$\ will be  meaningful. Fig. \ref{fig:omegas} shows the expectations for a 400-strong quasar mock sample simulated with uniform rms = 0.3 over the redshift range 0.2 -- 3.0 (details are provided in Marziani \& Sulentic, submitted). Quasars will yield  similar constraints on $ \Omega_\mathrm{M}$\ as supernov\ae\ since they are able to chart cosmic epochs when matter dominated cosmic expansion. In both cases, only statistical errors were considered. 

We are still   far   from the precision of absolute magnitude of type Ia supernov\ae\ whose dispersion is now estimated $\approx 0.15$ mag \citep{folatellietal10,hook13}. The dispersion in Eddington candles is also much  larger than the dispersion in the $r_\mathrm{BLR}$ -- $L$ relation, $\approx 0.13$ dex that has been  proposed by \citet{watsonetal11} as a valuable cosmological ruler.  Even if the dispersion in bolometric luminosity of the Eddington standard candles is disarmingly large (but it is likely that the rms can be significantly reduced  as it happened for supernov\ae; \citealt[][and references therein]{jacobyetal92}), our preliminary analysis has shown that they might be useful  if one exploits the possibility  of building large samples uniformly covering a wide redshift range.


\subsection{Caveats}

The physics behind xA sources is still poorly understood.  xA sources   have low equivalent width in {\em all} the strongest broad  emission lines i.e, C{\sc iv}$\lambda$1549 and  Balmer lines. The reason of the low equivalent width is not fully clear \citep{diamond-stanicetal09,shemmeretal10} but extreme Eddington ratio sources may have stripped the BLR of the lower column density gas \citep{netzermarziani10} leaving only the densest part of the BLR for low-ionization line emission \citep{marzianietal10,negreteetal13}.  
 A4 sources show extreme iron and aluminium emission, and this might be associated to chemical enrichment \citep{nagaoetal06b,juarezetal09}. Selective aluminium and silicon abundance enhancement can occur in supernova ejecta \citep{woosleyweaver95}, and such enhancement may be needed to explain in detail the emission line ratios observed in A3 and especially A4 sources \citep{negreteetal12}. This in turns lead us to the issue of a particular evolutionary state that may be associated with the extreme Eddington radiators, and with the feedback relation between star formation and black hole accretion \citep[e. g.,][]{silk13}. An important  open issue is the relation between extreme Eddington ratio and enhancement of circumnuclear star formation \citep{sanietal10}. 
 It is also not clear whether/how slim disk models  that predict a large $\Gamma_\mathrm{hard}$\ can consistently account for the soft X-ray excess i.e., for the large $\Gamma_\mathrm{soft}$. The asymptotic value of the bolometric luminosity of the slim disk also depends on the poorly known viscosity and opacity properties of the disk  \citep[e.g.,][and references therein]{dotanshaviv11}. 

We cannot ignore that there are caveats also for the radius-luminosity method: using $c \tau$\  as a standard linear ruler is certainly appealing, but is there here any well-known structure here that is employed as a standard ruler for cosmology? Since the broad line region is not resolved, it is not obvious if this is the case.  It is known since long \citep[e.g., ][]{netzer90} that  $r_\mathrm{BLR}$\ derived with reverberation mapping  can differ from what is ideally expected to measure i.e., an emissivity weighted distance from the central continuum source. The issue is certainly not settled  \citep{devereux13}.  Unclear aspects involve the uncertain effect of radiation forces and  the physical conditions that are likely to be different from object to objects, as pointed out by \citet{watsonetal11}. Low-$z$ reverberation measures are mainly for H$\beta$, but  H$\beta$\ observations in the IR are still difficult at the time of writing.  Resorting to prominent UV resonance lines will require a careful  inter calibration not unlike the one needed for combining optical and UV data for Eddington standard candles.



\section{Conclusion}

The potential of using quasars for cosmographic studies has not yet been exploited.  Promising methods based on quasar intrinsic properties involve the identification of ``Eddington standard candles'' and the use of the broad line region distance from the quasar continuum sources as a geometric ruler whose dependence on luminosity is known. Both methods are still relatively untested at present and should be further investigated in the near future.   Ideally, standard candles distributed uniformly over a wide redshift range (0 $\lesssim z \lesssim 4$) could even provide $z$-dependent constraint  on the physics of accelerated expansion, and specifically on the cosmic evolution of the dark energy equation of state \citep[e.g.][]{riessetal07}. 

The 4DE1 formalism appears to be the most promising path towards identifying extreme Eddington radiators. It remains to be seen whether   well-defined observational properties described in this paper and by \citet{wangetal13} could be exploited for refining a cosmologically useful sample. Eddington standard candles can already explore a range of distance where the metric of the Universe has not yet been ``charted''. A preliminary sample gives encouraging results but known systematic effects have to be carefully quantified and statistical errors reduced before a successful determination of at least $\Omega_\mathrm{M}$\ can be claimed. 


\paragraph{Acknowledgements} Funding for the SDSS and SDSS-II has been provided by the Alfred P. Sloan Foundation, the Participating Institutions, the National Science Foundation, the U.S. Department of Energy, the National Aeronautics and Space Administration, the Japanese Monbukagakusho, the Max Planck Society, and the Higher Education Funding Council for England. The SDSS Web Site is http://www.sdss.org/.

\bibliographystyle{model1b-num-names} 


\end{document}